\documentclass{elsart}
\usepackage{graphicx}
\usepackage{dcolumn}
\usepackage{bm}

\begin{document}


\begin{frontmatter}

\title{Scalar-Top Masses from  SUSY Loops \\
with  125 GeV $m_h$ and  precise $M_W$,$m_t$}


\author{$^1$Vernon Barger,\  $^1$Peisi Huang,\
        $^2$Muneyuki Ishida,\ $^3$Wai-Yee Keung\quad\quad}
\address{
$^1$Department of Physics, University of Wisconsin, Madison, WI 53706, USA
\\
$^2$Department of Physics, Meisei University, Hino, Tokyo 191-8506, Japan
\\
$^3$Department of Physics, University of Illinois at Chicago, IL 60607, USA
}

\date{\today}
\begin{abstract}
We constrain the masses of scalar-tops (stop) by analyzing the new
precision Tevatron measurement of the $W$-boson mass and the
LHC/Tevatron indications of a Higgs boson of mass 125.5$\pm$1 GeV. Our
study adopts Natural SUSY with low fine-tuning, which has multi-TeV
first and second generation squarks and a light Higgsino mixing
parameter $\mu$ =150 GeV.  An effective Lagrangian calculation is made
of $m_h$ to 3-loops using the H3m program with weak scale SUSY
parameters obtained from RGE evolution from the GUT scale in the
Natural SUSY scenario. The SUSY radiative corrections to the Higgs
mass imply maximal off-diagonal elements of the stop mass-matrix and a
mass splitting of the two stops larger than $400$~GeV.
\end{abstract}

\begin{keyword}
Higgs boson \sep top squark \sep SUSY
\PACS 14.80.Ly \sep 12.60.Jy
\end{keyword}
\end{frontmatter}



Supersymmetry(SUSY) is a theoretically attractive extension of the Standard Model(SM) that may explain the hierarchy of the weak scale and the Planck scale.  Of the SUSY particles, the lighter scalar top squark may
have a sub-TeV mass and be detectable by LHC experiments.
Existence of a light top-squark is particularly suggested by
the Natural SUSY
model\cite{Nold0,Nold1,Nold2,Nold3,Nold4,Nold5,Nold6,Nold7,Nold8,Nnew1,Nnew2,Nnew3,Nnew4,Nnew5,Nnew6,Nnew7,BBPM,RNSBaer,Gh,Feng,RR},
that has less fine tuning.
The first and second generation squarks have multi-TeV masses to mitigate unwanted
flavor changing neutral currents (FCNC) and large CP violation.
For a third generation scalar GUT-scale mass $m_0(3)<1$~TeV,
$m_{\tilde t_1}$ is less than 400~GeV from the running of
the RGE equations \cite{BBPM}.

A light top squark can give a significant radiative contribution to the $W$-boson mass.
The precision of $M_W$ has been improved  by recent Tevatron measurements;
$M_W=80,387\pm 12(stat.)\pm 15(syst.)$~MeV by the CDF collaboration\cite{CDFMW}
and $M_W=80,367\pm 13(stat.)\pm 22(syst.)$~MeV by the D0 collaboration\cite{D0MW}.
Including these measurements, the world average $M_W$ is shifted downward from \cite{TevMW}
$
M_W^{\rm exp} = 80,399\pm 26~{\rm MeV}$ to $80,385\pm 15~{\rm MeV}.
$
The SM prediction\cite{Sirlin,Marciano} of $M_W$ at 2-loop order is
\begin{eqnarray}
M_W^{\rm SM} &=& 80,361\pm 7~{\rm MeV}.
\label{eq1}
\end{eqnarray}
where we have used the numerical formula of  ref.\cite{MWSM} with central values of parameters\cite{parameters}
The uncertainties of the SM prediction of $M_W$ resulting from 
the uncertainties of these input parameters are summarized in Table~\ref{tab1}.

The LHC experiments have reported indications of a Higgs boson at mass
$125.3\pm 0.4_{\rm stat}\pm 0.5_{\rm syst}$ GeV in CMS data\cite{CMS} and at
$126.0\pm 0.4_{\rm stat}\pm 0.4_{\rm syst}$ GeV in ATLAS
data\cite{ATLAS}. Accordingly, we assume a Higgs boson mass of
$125.5\pm 1.$ GeV in our study. Then, the difference of the
experimental and SM values of $M_W$ is
\begin{eqnarray}
M_W^{\rm exp}-M_W^{\rm SM} &=& 24 \pm 15~{\rm MeV}.
\label{eq2}
\end{eqnarray}

\begin{table}[t]
\begin{center}
\begin{tabular}{lc}
       & \ \ \  $\delta M_W$\\
\hline
$\delta m_h=1.0$~GeV & \ \ \  $-0.5$~MeV\\
$\delta m_t=1$~GeV & \ \ \  $6.0$~MeV\\
$\delta M_Z=2.1$~MeV & \ \ \  $2.6$~MeV\\
$\delta (\Delta\alpha_{\rm had}^{(5)})=0.6\times 10^{-4}$ & \ \ \  $-1.1$~MeV\\
$\delta \alpha_s(M_Z)=0.0007$ & \ \ \  $-0.4$~MeV
\end{tabular}
\end{center}
\caption{Uncertainty of the SM $M_W$ prediction from the uncertainties of the parameters.
Beside these errors, there is another uncertainty due to missing higher order corrections,
which is estimated as about 4 MeV.\cite{MWSM}}
\label{tab1}
\end{table}
As can be seen in Table~\ref{tab1},
the largest source uncertainty in $M_W^{\rm SM}$ (of 6.0~MeV) is from the uncertainty
$\delta m_t=1$~GeV in the top mass measurement.
It is significantly smaller than the experimental uncertainty in
$M_W^{\rm exp}$ (of 15 MeV),
given in Eq.~(\ref{eq2}).

The contributions of SUSY particles to the one-loop calculation of $M_W$\cite{HeineHollik} along with the $W$ self-energy at the two loop level\cite{DjouHollik}
can account for the  $1.6\sigma$ deviation of the experimental value
from the SM prediction\cite{HeineHollik}. Conversely,
the $M_W$ measurement gives a constraint on the squark masses of the third generation,
$m_{\tilde t_1}$, $m_{\tilde t_2}$, and $m_{\tilde b_L}$.
We assume no mixing in sbottom sector since that off-diagonal element is proportional to
$m_b$; $m_{\tilde b_R}$ is irrelevant to $\delta M_W$.

The dominant SUSY radiative corrections to $m_h$ are due to loops
of $\tilde t_1$ and $\tilde t_2$. Implications of a 125 GeV Higgs boson for
supersymmetric models are investigated in ref.\cite{Djou}.
If $m_h$ is confirmed with the value of the present Higgs-boson signal $\sim$125.5~GeV,
the values of $m_{\tilde t_1},m_{\tilde t_2}$ and the top squark mixing angle
$\theta_{\tilde t}$ can be constrained from the measured $m_h$.
We investigate how a Higgs mass
$m_h = 125.5\pm 1.0 $~GeV and the new experimental value of $M_W$ constrain the third generation SUSY scalar top masses.\\

\noindent\underline{\it Constraint from $M_W$}\ \ \ \
The $M_W$ prediction is obtained by calculating the muon lifetime\cite{Sirlin,Marciano,HeineHollik}.
The SUSY correction $\Delta r$ to the Fermi constant $G_\mu$ is
\begin{eqnarray}
\frac{G_\mu}{\sqrt 2} &=& \frac{e^2}{8 s_W^2 M_W^2}(1+\Delta r)
\label{eq3}
\end{eqnarray}
where $s_W={\rm sin}\theta_W$ and $\theta_W$ is weak mixing angle which is defined by the experimental values of $W/Z$ pole mass $M_{W/Z}$ as
\begin{eqnarray}
c_W^2 &\equiv& {\rm cos}^2\theta_W =\frac{M_W^2}{M_Z^2}\ .
\label{eq4}
\end{eqnarray}
$\Delta r$ is calculated\cite{HeineHollik} in the MSSM,
and the corresponding $M_W$ prediction is obtained by iterative solution of the equation
\begin{eqnarray}
M_W^2 &=& M_Z^2\times\left\{ \frac{1}{2}+\sqrt{\frac{1}{4}-\frac{\pi\alpha}{\sqrt 2 G_\mu M_Z^2}[
1+\Delta r(M_W,M_Z,m_t,\cdots)]} \right\} .
\label{eq5}
\end{eqnarray}
Then, the correction to $M_W^2$ is at one-loop level is
\begin{eqnarray}
\delta M_W^2 &=& -M_Z^2\frac{c_W^2 s_W^2}{c_W^2-s_W^2}\Delta r\ .
\label{eq6}
\end{eqnarray}
$\Delta r$ is given by\cite{Sirlin,Marciano}
\begin{eqnarray}
\Delta r &=&
\frac{c_W^2}{s_W^2}\left(\frac{\delta M_Z^2}{M_Z^2}
-\frac{\delta M_W^2}{M_W^2}\right) +\Delta\alpha +(\Delta r)_{\rm rem.}
\label{eq7}
\end{eqnarray}
The first term on the left-hand-side
is the on-shell self-energy correction to gauge boson masses; $\frac{\delta M_Z^2}{M_Z^2}-\frac{\delta M_W^2}{M_W^2}=
-\frac{\Sigma^Z (M_Z^2)}{M_Z^2}+\frac{\Sigma^W (M_W^2)}{M_W^2}$ .
$\Delta \alpha$ is the radiative correction to the fine structure constant $\alpha$ .
The remainder term $(\Delta r)_{\rm rem.}$ includes
vertex corrections and box diagrams at one loop level which give subleading contributions
compared with the 1st term of Eq.~(\ref{eq7})\cite{HeineHollik}.

The main contribution to $\delta M_W$ is the on-shell gauge-boson
self energy,
which is well approximated\cite{DjouHollik,Heine2} with its value at zero momenta as
\begin{eqnarray}
\Delta r &\simeq&
-\frac{c_W^2}{s_W^2}\left(\frac{\Sigma^Z (0)}{M_Z^2}-\frac{\Sigma^W (0)}{M_W^2}\right)
= -\frac{c_W^2}{s_W^2}  \Delta \rho
\label{eq8}
\end{eqnarray}
where $\Delta \rho$ is the deviation of the $\rho$ parameter due to new physics in the EW
precision measurements. It is related to the $T$ parameter\cite{PeskinTakeuchi} by
\begin{eqnarray}
\Delta \rho &\simeq& \alpha(M_Z) T\ .
\label{eq9}
\end{eqnarray}

The squark, slepton, and neutralino/chargino loops contribute to $\Delta \rho$
at 1-loop level, which we denote as $\Delta \rho_0$.
The neutralino/chargino contributions are
small\cite{Hagiwara}, and the slepton contributions are suppressed relative to squark contributions by color,
and thus the squark contributions are dominant.
It is well known\cite{Veltman} that the weak $SU(2)_L$ isospin violation from
SUSY doublet masses gives non-zero
contributions to $\delta M_W$. The scalar-top sector is expected to have a
large $L-R$ mixing since the off-diagonal elements of the top squark mass matrix are proportional to $m_t$.
Finally, $\delta M_W$ is given by\cite{DjouHollik,Heine2}
\begin{eqnarray}
\delta M_W &\simeq& \frac{M_W}{2}\frac{c_W^2}{c_W^2-s_W^2}\Delta\rho_0,\nonumber\\
\Delta\rho_0 &=& \frac{3G_F}{8\sqrt 2 \pi^2}
[-s_{\tilde t}^2c_{\tilde t}^2 F_0(m_{\tilde t_1}^2,m_{\tilde t_2}^2)
 + c_{\tilde t}^2 F_0(m_{\tilde t_1}^2,m_{\tilde b_L}^2)
 + s_{\tilde t}^2 F_0(m_{\tilde t_2}^2,m_{\tilde b_L}^2) ]
\label{eq10}
\end{eqnarray}
where $F_0(a,b) \equiv a+b-\frac{2ab}{a-b}{\rm ln}\frac{a}{b}$ .
$s_{\tilde t}={\rm sin}\theta_{\tilde t}$, $c_{\tilde t}={\rm cos}\theta_{\tilde t}$,
and $\theta_{\tilde t}$ is the top squark mixing angle.
The 2-loop gluon$/$gluino exchange effects,
$\Delta\rho_{1,{\rm gluon}/{\rm gluino}}^{\rm SUSY}$, are neglected since
they are subleading compared with the 1-loop $\Delta\rho$ for
$M_{\rm susy} \stackrel{>}{\sim}$ 300~GeV\cite{Heine2}.
The prediction of $M_W$ in SUSY is then $M_W = M_W^{\rm SM} + \delta M_W$.
From Eq.~(\ref{eq10}) the $\delta M_W$ of Eq.~(\ref{eq2}) corresponds to
\begin{eqnarray}
\Delta \rho &=& (4.2 \pm 2.7)\times 10^{-4},\ \ \ \ T = 0.054\pm 0.034
\label{eq11}
\end{eqnarray}
The uncertainty is substantially reduced from that of the previous global
electroweak precision analyses:
$\Delta \rho =(3.67\pm 8.82)\times 10^{-4}$\cite{Lee},
$T=0.03\pm 0.11$\cite{Langacker}.

\begin{figure}[htb]
\begin{center}
\resizebox{0.7\textwidth}{!}{
 \includegraphics{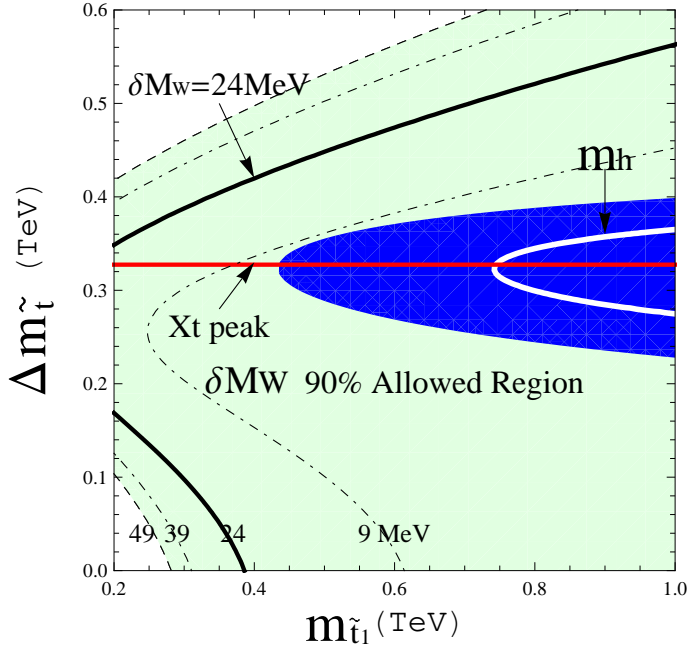}
}
\end{center}
\caption{Allowed regions
in the  $(m_{\tilde t_1},\Delta m_{\tilde t})$ plane for
$\theta_{\tilde t}=\frac{\pi}{4}$;
$\Delta m_{\tilde t}=(m_{\tilde t_2}-m_{\tilde t_1})$.
Black (Red) solid lines are
$\delta M_W=24$~MeV (maximum $m_h$ with $X_{t\ {\rm peak}}=-\sqrt{6}M_{\rm susy}$).
The Blue (dark-shaded) region is $m_h=123.5$ to $127.5$~GeV
and the white line represents its central value $m_h=125.5$~GeV.
The Green (medium-shaded) region is allowed by $\delta M_W$ at
90\% CL, and the dotdashed lines represent its 1$\sigma$ deviation, $\delta M_W=24\pm 15$~MeV.
}

\label{fig1}
\end{figure}

By using Eq.~(\ref{eq10}) with (\ref{eq2}), we can determine the allowed region
in the $m_{\tilde t_1},\Delta m_{\tilde t}$ plane
for a given value of $\theta_{\tilde t}$.  Here $\Delta m_{\tilde t}=(m_{\tilde t_2}-m_{\tilde t_1})$.
The case $\theta_{\tilde t}=\frac{\pi}{4}$ is shown in Fig.~\ref{fig1}.  Note that
$X_t$ and $\theta_{\tilde t}$ are independent because the soft-SUSY parameters in
the diagonal elements are different.

We also note that $m_{\tilde b_L}$ in Eq.~(\ref{eq10}) is given by
$m_{\tilde t_1}$, $m_{\tilde t_2}$, and $\theta_{\tilde t}$
\begin{eqnarray}
m_{\tilde b_L}^2
&=& m_{\tilde t_1}^2{\rm cos}^2\theta_{\tilde t}
+m_{\tilde t_2}^2{\rm sin}^2\theta_{\tilde t}
-m_t^2+m_b^2 -M_W^2{\rm cos}2\beta \ .
\label{eq12}
\end{eqnarray}
Equation (12) is symmetric under the exchange
\begin{eqnarray}
m_{\tilde t_1} \leftrightarrow& m_{\tilde t_2}\ ,\quad
c_{\tilde t} \leftrightarrow& s_{\tilde t},\ \ \hbox{ i.e. }
\theta_{\tilde t}\rightarrow \pi/2-\theta_{\tilde t}.
\label{eq13}
\end{eqnarray}
%

\noindent\underline{\it Constraint from $m_{h^0}$}\ \ \ \
The mass of the Higgs boson in the MSSM receives substantial radiative corrections to the tree level result. The scalar-top sector gives the dominant contribution,
for which $\Delta m_h^2 \propto m_t^4 /v^2$.
Tremendous efforts\cite{[5]}-\cite{Zhang:1998bm} have been expended to calculate $m_h$ with sufficient accuracy to compare with
LHC measurements, and
the Higgs mass has been calculated through the 3-loop level, $\alpha_t
\alpha_s^2$, for the leading $(m_t)^4$ corrections\cite{Kant1,Kant}
and partially at 4-loop level\cite{Martin}.  The dominant
contributions arise from supersymmetric loops involving the top
squarks, along with gluon and gluino exchanges.

There are several different approaches that have been used in the
theoretical evaluation of $m_h$ : perturbative calculation of the
Higgs self energy diagrams to (i) 2-loop and (ii) 3-loop orders, (iii)
effective field theory (EFT) methods based on second derivatives of an
effective Higgs potential, (iv) effective potential method based on
RGE evolution from the GUT scale, and (v) the effective Lagrangian
method. We succinctly summarize the five methodologies:

{\bf i)} The FeynHiggs package\cite{[29]} calculates $m_h$ diagrammatically in 2-loop order in the
on-shell(OS) renormalization scheme.

{\bf ii)} A MATHEMATICA program, H3m\cite{Kant}, does the three-loop calculation;
it is interfaced with the 2-loop FeynHiggs program for $m_h$ predictions.
A numerical 3-loop accuracy on $m_h$ has been estimated to be $<$ 1 GeV.  However,
its expansion in mass-squared ratios does not apply in some parameter regions relevant to
Natural SUSY.

{\bf iii)} In the EFT 2-loop leading-log approximation\cite{Esp,HH,Carena},
$m_h^2$ is calculated in the limit of stop matrix elements
$M_L=M_R$\cite{Carena,Carena1,Heinemeyer,Hollik,Zhang:1998bm},
and $M_L\gg M_R$\cite{Espinosa2}.

The $m_h^2$ formula in the general case with $M_L\neq M_R$, is given 
in the large $m_A$ limit by\cite{Carena}
\begin{eqnarray}
m_{h,{\rm EFT2}}^2(m_{\tilde t_1},m_{\tilde t_2},x_t) &=& M_Z^2 c_{2\beta}^2 + \frac{3 \bar m_t^4}{2\pi^2 v^2}\left[
\frac{1}{2}\tilde X_t + t \right.\nonumber\\
+\frac{1}{16\pi^2}
   \left( \frac{3\bar m_t^2}{v^2}-32\pi\alpha_s(\bar m_t) \right) &\times& \left.
\left( \tilde X_t t_{\rm max}+ \frac{t_{\rm max}^2+t_{\rm min}^2}{2}
   +(2t-t_{\rm max}-t_{\rm min})t_{\rm max}
  \right) \right]
\nonumber\\
t &\equiv& {\rm ln}\frac{m_{\tilde t_1}m_{\tilde t_2}}{\bar m_t^2},\quad
t_{\rm max}\equiv {\rm ln}\frac{M_{\rm max}^2}{\bar m_t^2},\quad
t_{\rm min}\equiv {\rm ln}\frac{M_{\rm min}^2}{\bar m_t^2},
\label{eq14}
\end{eqnarray}
where
$v\equiv 1/\sqrt{\sqrt 2 G_F}\simeq 246$~GeV
and the contribution from the sbottom sector can be omitted
so long as $\tan\beta$ is not close to its upper bound of $\sim$ 60.

In the above equation, $\tilde X_t$ is related with the
stop-mixing parameter $X_t=A_t-\mu {\rm cot}\beta$ by
\begin{eqnarray}
\tilde X_t &\equiv &
2|X_t|^2\frac{\ln(m_{\tilde t_2}^2/m_{\tilde t_1}^2)}
              {m_{\tilde t_2}^2-m_{\tilde t_1}^2}
+ |X_t|^4
\frac{2-\frac{m_{\tilde t_2}^2+m_{\tilde t_1}^2}
                    {m_{\tilde t_2}^2-m_{\tilde t_1}^2}
                        \ln(m_{\tilde t_2}^2/m_{\tilde t_1}^2) }
{(m_{\tilde t_2}^2-m_{\tilde t_1}^2)^2}
\ .
\label{eq17}
\end{eqnarray}
In Eqs.~(\ref{eq14}) and (\ref{eq17}) 
the $X_t$ is a quantity regularized with the renormalization scale 
$\mu=M_{\rm susy}$ in the $\overline{\rm MS}$ scheme,
while the running top quark mass $\bar m_t$ is evaluated at $\mu =\bar m_t$ itself in the
$\overline{\rm MS}$ scheme. $\bar m_t(\mu )$ was calculated in $\overline{DR}$ scheme by
ref.\cite{mtb} and in $O(\alpha_s^4)$\cite{mtb2,mtb3}.
Its value in the $\overline{\rm MS}$ scheme is $\bar m_t=163.71\pm 0.95$~GeV\cite{Langacker}
which corresponds to the on-shell top quark mass $M_t=173.4\pm 1.0$~GeV.


The $\tilde X_t$ in Eq.~(\ref{eq17}) is well approximated as
\begin{eqnarray}
\tilde X_t &=&  2 x_t^2 - \frac{x_t^4}{6},\quad  x_t \equiv \frac{X_t}{M_{\rm susy}}
\label{eq18}
\end{eqnarray}
with the choice of SUSY breaking scale
\begin{eqnarray}
M_{\rm susy} &=& \frac{m_{\tilde t_1}+m_{\tilde t_2}}{2}\ .
\label{eq19}
\end{eqnarray}

The $m_{h,{\rm EFT2}}^2$ of Eq.~(\ref{eq14}) has its maximum at
 $|x_t|=|(x_t)_{\rm max}|=\sqrt 6$ or $|X_t|=|(X_t)_{\rm max}|=\sqrt 6 M_{\rm susy}$, for which
 $\tilde X_t=6$.
It is also a common feature of the analytic EFT formula at 1- and 2-loop
levels\cite{Carena,Carena1,Heinemeyer,Hollik}.
A region $|X_t|\stackrel{>}{\scriptstyle \sim}\sqrt{6}M_{\rm susy}$ is theoretically not
allowed from considerations of false vacuum of charge and color symmetry breaking\cite{BKK,CCB,CCB1,CCB2}.


$M_{\rm max,min}$
are related to the stop squared-mass matrix $M_{\tilde t}^2$ in
on-shell(OS) renormalization scheme as
\begin{eqnarray}
M_{\tilde t}^{2} & \equiv &
\left(
\begin{array}{cc} M_L^{2} & M_t X_t^{\rm OS}\\ M_tX_t^{\rm OS} & M_R^{2} \end{array}  \right)
= \left(
\begin{array}{cc} m_{\tilde t_1}^2c_{\tilde t}^2+m_{\tilde t_2}^2s_{\tilde t}^2
 & -(m_{\tilde t_2}^2-m_{\tilde t_1}^2)c_{\tilde t}s_{\tilde t}\\
-(m_{\tilde t_2}^2-m_{\tilde t_1}^2)c_{\tilde t}s_{\tilde t}
 & m_{\tilde t_1}^2s_{\tilde t}^2+m_{\tilde t_2}^2c_{\tilde t}^2 \end{array}  \right)
\label{eq15}\\
(M^{OS})^2_{max,min} &\equiv& {\rm max,min}\{ M_L^{2},M_R^{2} \}
= \frac{m_{\tilde t_2}^2+m_{\tilde t_1}^2}{2} \pm
\sqrt{\left(\frac{m_{\tilde t_2}^2-m_{\tilde t_1}^2}{2}\right)^2-(M_t X_t^{\rm OS})^2}\ .
\ \ \ \ \ \ \ \
\label{eq16}
\end{eqnarray}
Our sign convention of $X_t$ agrees with that used in ref.\cite{Hollik}. 
$X_t^{\rm OS}$ is the on-shell stop mass matrix parameter.
The relation between $M_{\rm susy}^{\rm OS}$ and $X_t^{\rm OS}$ in OS scheme and those in $\overline{MS}$  scheme
are given in \cite{Hollik}, see also \cite{[25]}.
Here we treat $M_{\rm max,min}^{{\rm OS}}$ as being equal to
$M_{\rm max,min}$ in Eq.~(\ref{eq14}) since the difference is small (less than 4\%)
for $M_{\rm susy}>1$~TeV.

In Eq.~(\ref{eq19}),
the r.h.s is given by the on-shell stop masses and thus, more precisely
Eq.~(\ref{eq19}) is  $M_{\rm susy}^{\rm OS}$.
Here we regard $M_{\rm susy}^{\rm OS}$ as being equal to $M_{\rm susy}$ 
in $\overline{MS}$ scheme since the difference is small.

On the other hand, $X_t$ affects a relatively large difference 
between $\overline {DR}$ and OS schemes.
Numerically, we define the ratio 
\begin{eqnarray}  
\kappa &=& (X_t)_{\rm max}/ (X_t^{\rm OS})_{\rm max} 
\label{eq16A} 
\end{eqnarray} 
which is about 1.2 from the formula relating $\overline{\rm MS}$ and OS schemes 
given\footnote{$X_t^{\overline{\rm MS}}=X_t^{\rm OS}+\frac{\alpha_s}{3\pi}M_{\rm susy}
\left[ 8-\frac{X_t^2}{M_{\rm susy}^2}+\frac{4X_t}{M_{\rm susy}}+\frac{3X_t}{M_{\rm susy}}{\rm ln}\frac{M_{\rm susy}^2}{\bar m_t^2} \right]$ in 1-loop level\cite{Hollik}
 where the renormalization prescription is not specified in $O(\alpha_s)$ term.
} 
in Carena et al.\cite{Hollik}.
Coincidentally, $\kappa \approx \sqrt{6}/2.0$.
We choose this form because the factor 
$\sqrt 6$ matches the $x_t$ value in the $\overline{\rm MS}$
scheme giving maximum $\tilde X_t$ of Eq.~(\ref{eq18}) which leads to
maximum $m_{h,EFT2}^2$ of Eq.~(\ref{eq14}).
The 2.0 in the denominator is given as a numerical value 
of the ratio $(X_t^{\rm OS})_{\rm max}/M_{\rm susy}$ in 
ref.\cite{Hollik}.
We have also checked the ratio (\ref{eq16A}) by using Isajet 7.83\cite{ISASUSY}:
Isajet adopts the $\overline{DR}$ scheme and
$\overline{DR}\simeq\overline{MS}$ and converts to OS stop masses using\cite{[25]}.
 Isajet outputs of $X_t^{\overline{DR}}$ and on-shell
stop masses numerically consistent with the relation
$(X_t^{\overline{\rm DR}})_{\rm max}/(X_t^{\rm OS})_{\rm max} =\sqrt 6/2.0$.
(See, also, the caption of Fig.\ref{fig4}.)
We apply this relation (\ref{eq16A}) in the region close to ``maximal mixing", 
$|X_t|/M_{\rm susy}\sim \sqrt 6$:
\begin{eqnarray}
X_t &=& \kappa X_t^{\rm OS},\ \ \ \kappa=\sqrt 6/2.0
\label{eq16B}
\end{eqnarray}


The EFT method is not gauge-fixing invariant\cite{Martin}.
Nonetheless, it is found to give a good approximation when compared to other methods.
The formulae (\ref{eq14}) with (\ref{eq17}) gives larger $m_h$ values by about 1 GeV
than the results of H3m with the inputs of the natural SUSY benchmark points,
as will be commented on below.

The $m_h^2$ formula obtained from the 2-loop diagrammatic approach (i) can be matched
to the EFT formula above by adjusting the renormalization prescription\cite{Hollik},
except for additional non-logarithmic terms in the diagrammatic formula that give
asymmetric heights of the peak $m_h$ at $X_t>0$ and $X_t<0$.
The latter contributions arise from SUSY threshold effects 
that are not taken into account in the RGE running down from the SUSY-breaking scale
 that includes logarithms of $M_{\rm susy}/ \bar m_t$. 

{\bf iv)} In the unification approach, RGEs are evolved from the GUT coupling unification scale\cite{mtb}, where the 1st and 2nd  generation scalars in Natural SUSY have a $m_0 \sim 10$ TeV mass
and the 3rd-generation scalars have $m_0\sim  1$ TeV\cite{BBPM,BB2}. The Higgs potential at the SUSY breaking scale $M_{\rm susy}$ is based on one-loop MSSM radiative corrections that are RGE improved.
With the choice of $M_{\rm susy} = \sqrt{m_{\tilde t_1}m_{\tilde t_2}}$, the
most important two-loop effects\cite{HHH} are included in the effective
potential.  The RGE evolution is implemented with the
ISASUSY package\cite{ISASUSY,ISASUSYA}, with a scan over GUT scale parameters.

{\bf v)} In the effective Lagrangian approach, the gauge couplings, the Yukawa couplings,
and the soft-SUSY terms are also RGE evolved to the weak scale from high scale boundary values,
 where the gauge couplings unify.
The ISASUSY program for this RGE evolution incorporates
SUSY threshold effects.\cite{ISASUSY,ISASUSYA}.
The weak scale parameters so obtained are taken as input to the diagrammatic calculation
at 2-loop order by the FeynHiggs\cite{[29]} or 3-loop order by the H3m\cite{Kant}.
It has been argued\cite{Martin} that this method may provide 
the most accurate evaluation of
the leading and next-to-leading contributions to $m_h$ in 3-loop order
in the approximation of large QCD and top-quark Yukawa couplings.

We adopt the latter approach in the framework of natural SUSY using
ISASUSY\cite{ISASUSY,ISASUSYA}, with a scan over GUT scale input parameters.
We have also corrected the sign convention of $X_t$ in ISASUSY in order to match ours.
We then
evaluate $m_h$ using the H3m program with the ISASUSY input for the SUSY
parameters at the weak scale.
Specifically, we adopt the benchmark
line NS3 of Ref.\cite{BBPM} that has a Higgsino mass term $\mu =150$~GeV and other Natural SUSY benchmark points RNS1 and RNS2 of Ref \cite{RNSBaer}.
\footnotemark.
\footnotetext[1]{The SOFTSUSY\cite{softsusy}, SPheno\cite{spheno,spheno2} and SuSpect\cite{suspect} codes use the same algorithm as Isajet\cite{ISASUSY,ISASUSYA}  and employ similar threshold transitions matching the MSSM to the SM. The four codes produce mass spectrum in the mSUGRA model that are in close agreement. The Isajet\cite{ISASUSY,ISASUSYA}code provides the NUHM2 model of our interest. }
The NS3 gives $m_h=123.5$~GeV that is consistent with the LHC experimental value.
There is a strong preference for $A_t(M_{\rm susy})> 0$ and
tan$\beta > 10$ in Natural SUSY\cite{BBPM}.
\begin{figure}[htb]
\begin{center}
\resizebox{0.8\textwidth}{!}{
 \includegraphics{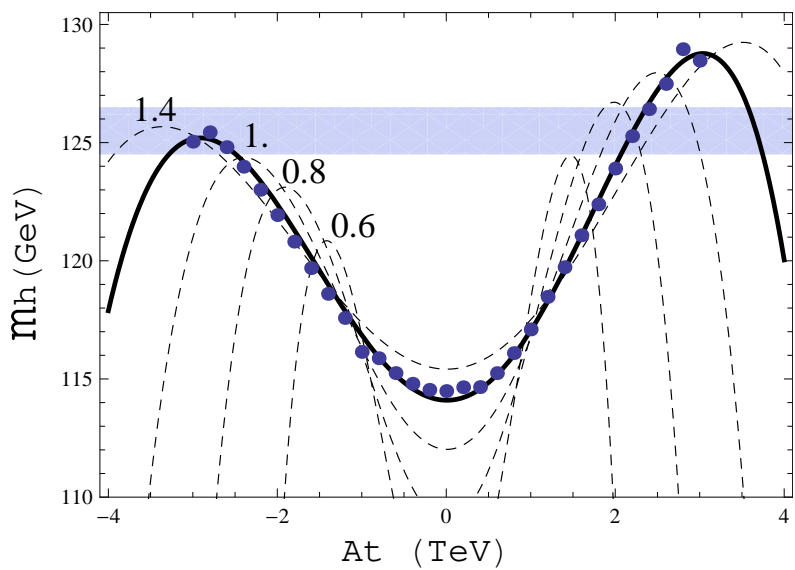}
}
\end{center}
\caption{$A_t(M_{\rm susy})$ dependence of $m_h$ in 3-loop calculation by H3m
with the effective Lagrangian method. (Solid circles).
The input parameters are a natural SUSY benchmark line (NS3):
$(m_{\tilde t_1,B},m_{\tilde t_2,B})=(812.5,1623.2)$~GeV which corresponds to
$M_{\rm susy}=1212.9$~GeV. It is
obtained by varying the third generation scalar mass $m_0$\cite{BBPM}
at the unification scale:
The solid line is the formula, Eq.~(\ref{eq20})
that is designed to numerically reproduce the effective Lagrangian result.
The dashed lines are obtained from the formula (\ref{eq21}) with inputs
$(m_{\tilde t_1},m_{\tilde t_2})=(m_{\tilde t_1,B}+\delta m,m_{\tilde t_2,B}+\delta m)$
with various $\delta m$ values corresponding to
$M_{\rm susy}(=\frac{m_{\tilde t_1}+m_{\tilde t_2}}{2})=0.6,0.8,1.0,1.4$~TeV.
$m_h=125.5\pm 1$~GeV is shown by blue band.
}
\label{fig2}
\end{figure}
Since $\mu$ is small in natural SUSY, $X_t$ is approximately $A_t$ for
$A_t \sim $ TeV.
We should note that variations of the masses of the 1st and 2nd generations and gauginos
from the NS3 inputs
have little effect on $m_h$ since they are heavy in Natural SUSY scenario.

The $m_h$ effective Lagrangian result with the NS3 input parameters
can be numerically represented by the formula
\begin{eqnarray}
m_h^2 &= & m_{h,B}^2(x_t) \equiv M_Z^2 c_{2\beta_B}^2
+ \frac{3\bar m_t^4}{2\pi^2 v^2}\left[ c_0+(c_1+c_2 x_t)\tilde X_t  \right] \nonumber\\
\tilde X_t &\equiv & 2x_t^2\left(1-\frac{x_t^2}{12}\right) ,
\ \ x_t\equiv \frac{X_t}{M_{\rm susy,B}}
\label{eq20}
\end{eqnarray}
where the subscript $B$ means the NS3 Benchmark point:
$c_{2\beta_B}={\rm cos}2\beta_B$ is calculated from tan$\beta_B=19.4$.
$M_{\rm susy,B}$ is the SUSY breaking scale corresponding to
$(m_{\tilde t_1,B},m_{\tilde t_2,B})=(812.5,1623.2)$~GeV;
$M_{\rm susy,B}=(812.5+ 1623.2)/2=1212.9$~GeV.
The coefficients \\
$$ (c_0,c_1,c_2)=(2.661,0.2874,0.01717)  $$
have been determined by a least-squares fit with some weighting of the maximal $m_h$ region.

We use $m_{h,B}$ of Eq.~(\ref{eq20}) as our benchmark
at a given value of $x_t$.
$m_h$ values with different $m_{\tilde t_{1,2}}$ and
$M_{\rm susy}=\frac{m_{\tilde t_1}+m_{\tilde t_2}}{2}$ inputs
are considered to be given with sufficient accuracy by shifting from
$m_{h,B}$ with a common value of $x_t$ through 2-loop analytic formula (\ref{eq14}).
\begin{eqnarray}
m_h^2(m_{\tilde t_1},m_{\tilde t_2},x_t,{\rm tan}\beta) = && \nonumber\\
m_{h,B}^2(x_t)
+ [m_{h,{\rm EFT2}}^2&&(m_{\tilde t_1},m_{\tilde t_2},x_t,{\rm tan}\beta)
 - m_{h,{\rm EFT2}}^2(m_{\tilde t_1,B},m_{\tilde t_2,B},x_t,{\rm tan}\beta_B)]
\ \ \ \ \ \ \ \ \label{eq21}
\end{eqnarray}
In order to estimate the intrinsic uncertainty, we also consider the other natural SUSY
benchmark points, RNS1 and RNS2\cite{BBPM}, where $m_h$ is estimated by using 
Isajet 7.83.
\begin{eqnarray}
\begin{array}{cc|c|c|c|c|c|c}
   & m_{\tilde t_1} & m_{\tilde t_2} & M_{\rm susy}
 & A_t^{\rm OS} & {\rm tan}\beta & m_h({\rm Isajet}) & m_h({\rm Eq.(\ref{eq21})})\\
{\rm RNS1} & 1416 & 3425 & 2420 & 3764 & 10 & 123.7 & 124.1\\
{\rm RNS2} & 1843 & 4921 & 3382 & 5054 & 8.55 & 125.0 & 123.4\\
\end{array}
\label{eq21A}
\end{eqnarray}
Here the masses and the $A_t^{\rm OS}$ are given in units of GeV.
The predictions from Eq.~(\ref{eq21}) are given in the final column.
Our formula (\ref{eq21}) is made by using a special input of NS3 benchmark point with $M_{\rm susy}\simeq 1.2$~TeV, but it can be applied to wide range of cases
with fairly good accuracy.
The theoretical error of Eq.~(\ref{eq21}) is conservatively considered to be 2 GeV in whole
range of parameters in natural SUSY scenario.

In order to see the $M_{\rm susy}$ dependence of $m_h$, we shift the $m_{\tilde t_{1,2}}$
from the NS3 benchmark values commonly with $\delta m$.
The results are shown by dashed lines in Fig.~\ref{fig2}, which suggests the necessity
of the maximal mixing condition when $X_t\simeq \sqrt 6 M_{\rm susy}$\cite{Stal,Heng}.
The peak value of $m_h$ gradually increases with $\sim$ln$M_{\rm susy}$.
The Higgs mass constraint $m_h>124.5$~GeV requires a SUSY
breaking scale $M_{\rm susy}\stackrel{>}{\scriptstyle \sim}0.6$~TeV.

\begin{figure}[htb]
\begin{center}
\resizebox{0.8\textwidth}{!}{
 \includegraphics{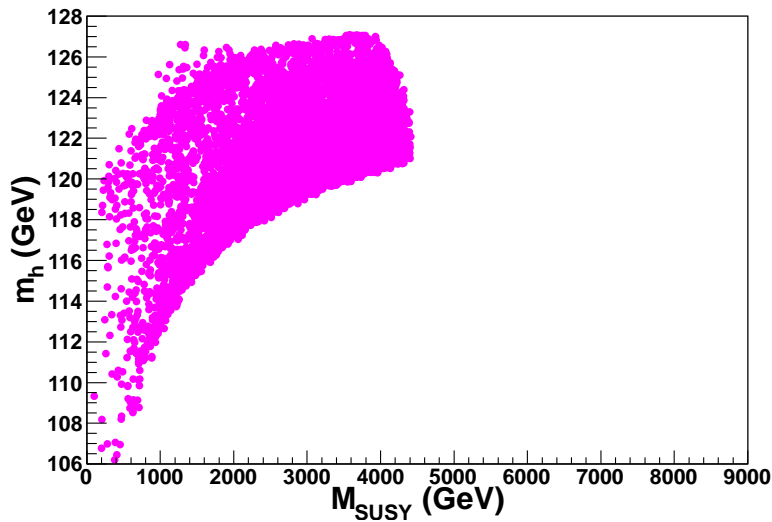}
}
\end{center}
\caption{$M_{\rm susy}$ dependence of $m_h$ in Natural SUSY points
following Ref.\cite{BBPM}.
The points are obtained from a scan over GUT scale parameters:
the common scalar mass of the first two generations
$m_0(1,2):5-50$ TeV, the third generation squark mass
$m_0(3):0-5$ TeV, the common gaugino mass $m_{1/2}:0-5$ TeV,
$-4< A_t /m_0(3) < 4$,
$m_A:0.15-2$ TeV, tan$\beta :1-60$. See, ref.\cite{BBPM}.
}
\label{fig3}
\end{figure}

\begin{figure}[htb]
\begin{center}
\resizebox{0.8\textwidth}{!}{
 \includegraphics{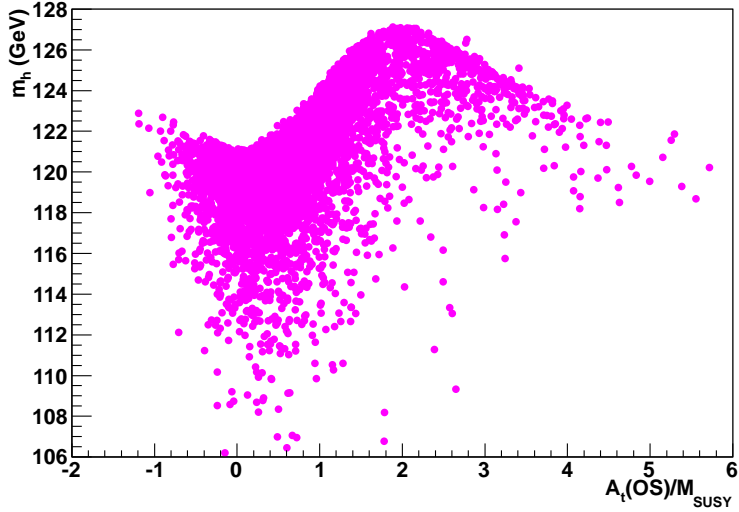}
}
\end{center}
\caption{$A_t(M_{\rm susy}) /M_{\rm susy}$ dependence of $m_h$ in natural SUSY scan points.
$X_t = A_t - \mu$cot$\beta\simeq A_t$ since $\mu$ is small, 150~GeV.
The maximum of $m_h$ is not obtained at $A_t(M_{\rm susy}) /M_{\rm susy}= \sqrt 6$ but at about $2$,
which is due to the difference of renormalization prescription of ISASUSY program, on-shell(OS)
renormalization, and the EFT approach using the $\overline{\rm MS}$ scheme. See, ref.\cite{Carena}.
}
\label{fig4}
\end{figure}

The $M_{\rm susy}$ dependence of $m_h$ in Natural SUSY points following Ref.\cite{BBPM} are shown
in Fig.~\ref{fig3}.
The points indicates a $\ln M_{\rm susy}$ dependence, and in order to explain $m_h>124.5$~GeV,
it is indeed plausible that $M_{\rm susy}>1$~TeV.

The maximal mixing condition $|X_t^{\rm OS}|\simeq 2 M_s$,
which corresponds to $|X_t|\simeq \sqrt 6 M_s$ 
in the $\overline{DR}$ or $\overline{MS}$ scheme,
can be obtained\cite{Badziak,BKK} by RGE running from the SUSY-GUT scale,
as illustrated for Natural SUSY in Fig.~\ref{fig4}; note that $A_t<0$ is almost absent.
The generated points are mainly in the region $0<A_t<2$;
however, although improbable from the scan, the maximal mixing $X_t=\sqrt 6 M_{\rm susy}$ is possible in Natural SUSY.

\begin{figure}[htb]
\begin{center}
\resizebox{0.8\textwidth}{!}{
 \includegraphics{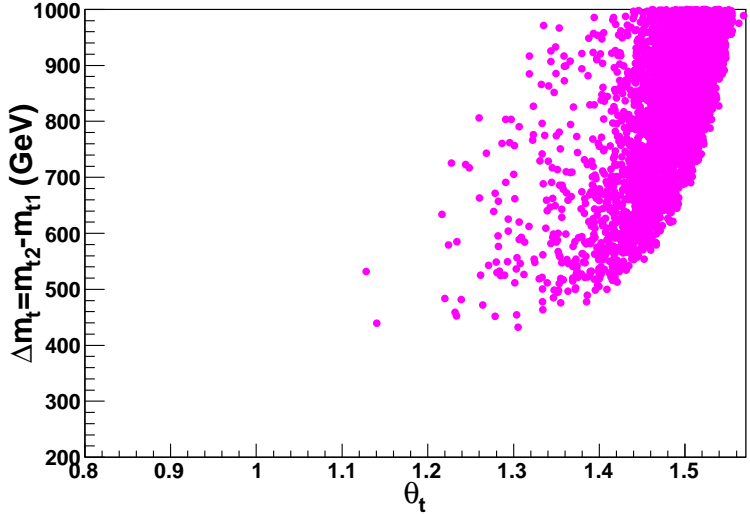}
}
\end{center}
\caption{$\theta_{\tilde t}$ dependence of stop mass difference 
$\Delta m_{\tilde t}=m_{\tilde t_2}-m_{\tilde t_1}$.
}
\label{fig5}
\end{figure}

By taking $m_h = 125.5\pm 2.~{\rm GeV}$
as a constraint to Eq.~(\ref{eq21}), we can determine the allowed region in
$(m_{\tilde t_1},m_{\tilde t_2})$ plane for a given value of $\theta_{\tilde t}$.
Here we allow a somewhat large uncertainty of $m_h$, 2~GeV,
because of the theoretical uncertainty of our formula (\ref{eq21}).
The Higgs mass constraint severely constrains the top
squark sector parameters, especially in that
$\Delta m_{\tilde t}(\equiv m_{\tilde t_2}-m_{\tilde t_1})$ has a lower limit.
From an Isajet scan over GUT scale parameters, 
we obtain the $\theta_{\tilde t}$ dependence of $\Delta m_{\tilde t}$ in Fig.~\ref{fig5}. 
Almost all data points have large $\theta_{\tilde t}$, $1.3<\theta_{\tilde t}<\frac{\pi}{2}$, which means $\tilde t_1\simeq \tilde t_R$.
$\Delta m_{\tilde t}$ decreases as
$\theta_{\tilde t}$ decreases from $\frac{\pi}{2}$.
Actually $\theta_{\tilde t}$ has a lower limit of 1.1 and 
we find that the on shell stop mass difference is bounded by
\begin{eqnarray}
\Delta m_{\tilde t}
\ge  400~{\rm GeV}.
\label{eq24}
\end{eqnarray}

\noindent\underline{\it Concluding Remarks}\ \ \ \
We have studied the implications for the scalar top sector of
the recent Tevatron $M_W$ measurements and the LHC and Tevatron indications of a 125 GeV Higgs boson.
We utilized the $H3m$ package to evaluate $m_h$ through 3-loops
in an effective Lagrangian approach with RGE evolution from the GUT scale.
Natural SUSY was assumed, for which the third generation scalar quarks are much lighter
than the multi-TeV masses of squarks of the first two generations and
the Higgsino mixing parameter $\mu$ is small, ~150 GeV.
A maximal Higgs mass is attained that is close to the LHC experimental indications.
The condition for maximal Higgs mass is an off-diagonal value of
the stop mixing matrix $X_t=\sqrt{6}M_s$\ in the $\overline{DR}$ renormalization
 scheme, which requires an on-shell soft-SUSY parameter
at the weak scale of $A_t(M_{\rm susy})\approx 2$ TeV.
The minimum value of the mass-splitting of two top
squark states was found to be 400 GeV.
As can be seen in Fig.~\ref{fig1}, the allowed region from the $m_h$ constraint
(blue region) satisfies the $M_W$ constraint at 90\% Confidence Level, 
independent of the value of $\theta_{\tilde t}$.
For $\theta_{\tilde t} = \frac{\pi}{4}$
a top-squark with sub-TeV mass  is somewhat favored by the $M_W$ data; 
$m_{\tilde t_1}<500$~GeV is possible for almost all values $\theta_{\tilde t}$ 
when $\tilde t_1\simeq \tilde t_R$.  Precise experimental determination of $m_h$ 
at the LHC will tighten the restrictions on the top squark masses.  
The detection of the scalar top states at the LHC would establish 
the SUSY theoretical underpinning of electroweak symmetry breaking.

V.B. thanks Howie Baer for valuable conversations on this topic.
M.I. is very grateful to the members of the Phenomenology Institute of University of Wisconsin-Madison for hospitality.
This work was supported in part by the U.S. Department of Energy under
grants No. DE-FG02-95ER40896 and DE-FG02-12ER41811,
in part by KAKENHI(2274015, Grant-in-Aid for Young Scientists(B)) and in part by Special Researcher grant of Meisei University.

\nocite{*}

\bibliography{apssamp}

\begin{thebibliography}{00}
\bibitem{Nold0} M.~ Dine, A.~Kagan and S.~Samuel, Phys. Lett. B 243 (1990) 250
\bibitem{Nold1}
S.~Kelley, J.~Lopez, D.~Nanopoulos, H.~Pois, and K.~Yuan, Nucl. Phys. B398, 3 (1993).
\bibitem{Nold2}
V.~Barger, M.~Berger, P.~Ohmann, Phys.Rev. D49 (1994) 4908-4930.
\bibitem{Nold3} S.~Dimopoulos and G.~Giudice, Phys.~Lett.~B{\bf 357}, 573 (1995). arXiv:hep-ph/9507282 [hep-ph].
\bibitem{Nold4} F.~Gabbiani, E.~Gabrielli, A.~Masiero, and L.~Silvestrini,
 Nucl.~Phys.~B{\bf 477}, 321 (1996) [arXiv:hep-ph/9604387 [hep-ph]].
\bibitem{Nold5}
A.~Cohen, D.~Kaplan, and A.~ Nelson Phys.Lett. B388 (1996) 588-598.
\bibitem{Nold6}
K.~Chan, U.~Chattopadhyay and P.~Nath, Phys. Rev. D 58 (1998) 096004.
\bibitem{Nold7}  J.~L.~Feng, K.~T.~Matchev, and T.~Moroi, Phys.~Rev.~ D{\bf 61} (2000) 075005 [arXiv:hep-ph/9909334 [hep-ph]]; Phys.~Rev.~Lett. {\bf 84}, 2322 (2000) [arXiv:hep-ph/9908309 [hep-ph]].
\bibitem{Nold8} R.~Kitano and Y.~Nomura, Phys.~Lett.~B{\bf 631}, 58 (2005) [arXiv:hep-ph/0509039 [hep-ph]].
%
\bibitem{Nnew1}
C.~Brust, A.~Katz, S.~Lawrence and R.~Sundrum, JHEP {\bf 1203} (2012) 103 
[arXiv:1110.6670 [hep-ph]].
\bibitem{Nnew2}
S.~Akula, M.~Liu, P.~Nath, and G.~Peim, Phys.~Lett.~B{\bf 709}, 192 (2012) [arXiv:1111.4589 [hep-ph]].
\bibitem{Nnew3}
R. Essig, E. Izaguirre, J. Kaplan and J. G. Wacker, JHEP {\bf 1201} (2012) 074 [arXiv:1110.6443 [hep-ph]].
\bibitem{Nnew4}
M. Papucci, J. T. Ruderman and A. Weiler, JHEP {\bf 1209} (2012) 035 
[arXiv:1110.6926 [hep-ph]].
\bibitem{Nnew5}
L.~Hall, D.~Pinner, and J.~Rudermn, JHEP {\bf 1204}, 131(2012)
\bibitem{Nnew6}
S.~King, M.~Muhlleitner and R.~Nevzorov, Nucl.Phys. B{\bf 860}, 207 (2012) 
[arXiv:1201.2671 [hep-ph]].
\bibitem{Nnew7} N.~Arkani-Hamed, talk at WG2 meeting, Oct. 31, 2012, CERN, Geneva.
%
\bibitem{BBPM}
H.~Baer, V.~Barger, P.~Huang, JHEP {\bf 1111}, 031 (2011) [arXiv:1107.5581 [hep-ph]].\\
%
H.~Baer, V.~Barger, P.~Huang, X.~Tata,in JHEP {\bf 1205}, 109 (2012) [arXiv:1203.5539[hep-ph]].\\
%
\bibitem{RNSBaer} H.~Baer, V.~Barger, P.~Huang, A.~Mustafayev, and X.~Tata, arxiv:1207.3343[hep-ph].
%
\bibitem{Gh} D.~M.~Ghilencea, H.~M.~Lee, and M.~Park, JHEP 1207 (2012) 046 
[arXiv:1203.0569 [hep-ph]].
\bibitem{Feng} J.~Feng and D.~Sanford, Phys.Rev. D86 (2012) 055015 
[arXiv:1205.2372 [hep-ph]].
\bibitem{RR} L.~Randall and M.~Reece, arXiv:1206.6540 [hep-ph].
%
\bibitem{CDFMW} CDF Collaboration, T. Aaltonen et al., Phys. Rev. Lett. {\bf 108} 151803 (2012) [arXiv:hep-ex:/1203.0275].
\bibitem{D0MW} D0 Collaboration, V.M. Abazov et al., Phys. Rev. Lett. {\bf 108} 151804 (2012) [arXiv:hepex:/1203.0293].
\bibitem{TevMW} Tevatron Electroweak Working Group, arXiv:1204.0042v2[hep-ex].
\bibitem{Sirlin} A.~Sirlin, Phys.~Rev.~D{\bf 22}, 971 (1980).
\bibitem{Marciano} W.~J.~Marciano and A.~Sirlin, Phys.~Rev.~D{\bf 22}, 2695 (1980); {\bf 31}, 213(E) (1985).
%
\bibitem{MWSM} M.~Awramik, M.~Czakon, A.~Freitas, and G.~Weiglein, Phys.~Rev.~D{\bf 69}, 053006 (2004)
\bibitem{parameters} $m_h = 125.5\pm 1.{\rm GeV}$,\ \ $m_t = 173.2\pm 0.6\pm 0.8~{\rm GeV}$,
\ \ $M_Z=91.1876\pm 0.0021~{\rm GeV}$ \\
$\alpha_s(M_Z)=0.1184\pm 0.0007$\ ,\
$\Delta\alpha_{\rm had}^{(5)} = (275.7\pm 0.6) \times 10^{-4}$.
$\Delta \alpha_{\rm had}^{(5)}$ is a correction to the fine structure constant from quark loops.
%
\bibitem{CMS} CMS Collaboration, arXiv:1207.7235[hep-ex].
\bibitem{ATLAS} ATLAS Collabotation, arXiv:1207.7214[hep-ex].
%
\bibitem{HeineHollik} S.~Heinemeyer, W.~Hollik, D.~Stockinger, A.~M.~Weber,
and G.~Weiglein, JHEP{\bf 08}, 052 (2006).
\bibitem{DjouHollik} A.~Djouadi, P.~Gambino, S.~Heinemeyer, and W.~Hollik, C.~Junger, and G.~Weiglein,
Phys.~Rev.~D{\bf 57}, 4179 (1998).
\bibitem{Djou} A.~Arbey, M.~Battaglia, A.~Djouadi, F.~Mahmoudi and J.~Quevillon,  Phys.~Lett. B{\bf 708} (2012) 162 
[arXiv:1112.3028[hep-ph]].
\bibitem{Heine2} S.~Heinemeyer, W.~Hollik, and G.~Weiglein, Phys.~Rep.~{\bf 425}, 265 (2006).
\bibitem{PeskinTakeuchi} M.~E.~Peskin and T.~Takeuchi, Phys.~Rev.~D{\bf 46}, 381 (1992).
\bibitem{Hagiwara} K.~Hagiwara, R.~Liao, A.D.~Martin, D.~Nomura, T.~Teubner,
J.~Phys.~G{\bf 38}, 085003 (2011), arXiv:1105.3149 [hep-ph] .
\bibitem{Veltman} M.~Veltman, Nucl.~Phys.~B{\bf 123}, 89 (1977).
\bibitem{Lee} H.~M.~Lee, V.~Sanz, M.~Trott, JHEP 1205 (2012) 139 
[arXiv:1204.0802[hep-ph]].
\bibitem{Langacker} J.~Erler and P.~Langacker, Particle Data Group,  J. Beringer et al., Phys. Rev. D86, 010001 (2012).
\bibitem{[5]} S.P. Li and M. Sher, Phys. Lett. B {\bf 140}, 339 (1984).
\bibitem{[6]} M.S. Berger, Phys. Rev. D {\bf 41}, 225 (1990).
\bibitem{[7]} H.E. Haber and R. Hempfling, Phys. Rev. Lett. {\bf 66}, 18159 (1991);
 Phys. Rev. D {\bf 48}, 4280(1993) [hep-ph/9307201].
\bibitem{HH} R. Hempfling and A.H. Hoang, Phys. Lett. B {\bf 331}, 99 (1994) [hep-ph/9401219].
\bibitem{[8]} Y. Okada, M. Yamaguchi and T. Yanagida, Prog. Theor.
Phys. {\bf 85}, 1 (1991); Phys. Lett. B {\bf 262}, 54 (1991).
\bibitem{[9]} J. Ellis, G. Ridolfi and F. Zwirner, Phys. Lett. B {\bf 257}, 83
(1991); Phys. Lett. B {\bf 262}, 477 (1991).
\bibitem{[10]} R. Barbieri, M. Frigeni and F. Caravaglios, Phys. Lett.
B {\bf 258}, 167 (1991).
\bibitem{[11]} A. Yamada, Phys. Lett. B {\bf 263}, 233 (1991); Z. Phys. C {\bf 61}, 247 (1994).
\bibitem{Esp} J.R. Espinosa and M. Quiros, Phys. Lett. B {\bf 266}, 389 (1991).
\bibitem{[12]}
J.A. Casas, J.R. Espinosa, M. Quiros and A. Riotto, Nucl. Phys. B 436, 3 (1995) [Erratum-ibid. B 439, 466(1995)] [hep-ph/9407389].\\
J.R. Espinosa and R.J. Zhang, JHEP 0003, 026 (2000)[hep-ph/9912236].
%
\bibitem{Espinosa2}
 J.~R.~Espinosa and R.~-J.~Zhang,
 Nucl.\ Phys.\ B {\bf 586}, 3 (2000)  [hep-ph/0003246].\\
J.~Espinosa and I.~Navarro,
 Nucl.~Phys.~B{\bf 615}, 82 (2001)[arXiv:hep-ph/0104047].
%
\bibitem{[13]} A. Brignole, Phys. Lett. B {\bf 281}, 284 (1992).\\
G. Degrassi, P. Slavich and F. Zwirner, Nucl. Phys. B
{\bf 611}, 403 (2001) [hep-ph/0105096].\\
A. Brignole, G. Degrassi, P. Slavich and F. Zwirner, Nucl.
Phys. B {\bf 631}, 195 (2002) [hep-ph/0112177]; Nucl. Phys.
B {\bf 643}, 79 (2002) [hep-ph/0206101].
\bibitem{[14]} M. Drees and M.M. Nojiri, Nucl. Phys. B 369, 54 (1992);
Phys. Rev. D 45, 2482 (1992).
\bibitem{[15]} K. Sasaki, M. Carena and C.E.M. Wagner, Nucl. Phys. B {\bf 381}, 66 (1992).\\
J. Kodaira, Y. Yasui and K. Sasaki, Phys. Rev. D {\bf 50}, 7035 (1994) [hep-ph/9311366].
\bibitem{[16]} P.H. Chankowski, S. Pokorski and J. Rosiek, Phys. Lett.
B {\bf 274}, 191 (1992); Nucl. Phys. B {\bf 423}, 437 (1994) [hep-ph/9303309].
\bibitem{[22]} A. Dabelstein, Z. Phys. C 67, 495 (1995) [hepph/
9409375].
\bibitem{[23]} M. Carena, J.R. Espinosa, M. Quiros and C.E.M. Wagner,
Phys. Lett. B {\bf 355}, 209 (1995) [hep-ph/9504316].\\
A. Pilaftsis and C.E.M. Wagner, Nucl. Phys. B {\bf 553}, 3 (1999) [hep-ph/9902371].\\
M. Carena, J.R. Ellis, A. Pilaftsis and C.E.M. Wagner,
Nucl. Phys. B {\bf 586}, 92 (2000) [hep-ph/0003180].\\
M. Carena, J.R. Ellis, A. Pilaftsis and C.E.M. Wagner,
Nucl. Phys. B {\bf 625}, 345 (2002) [hep-ph/0111245].
\bibitem{[25]} D.M. Pierce, J.A. Bagger, K.T. Matchev and R.J. Zhang,
Nucl. Phys. B {\bf 491}, 3 (1997) [hep-ph/9606211].
\bibitem{[29]} S. Heinemeyer, W. Hollik and G. Weiglein, Comput.
Phys. Commun. {\bf 124}, 76 (2000) [hep-ph/9812320].\\
M. Frank, S. Heinemeyer, W. Hollik and G. Weiglein, hep-ph/0202166.\\
T. Hahn, W. Hollik, S. Heinemeyer and G.Weiglein, hep-ph/0507009.\\
S. Heinemeyer, W. Hollik and G. Weiglein, Eur. Phys. J. C {\bf 9}, 343 (1999) [hep-ph/9812472].
S. Heinemeyer, Eur. Phys. J. C {\bf 22}, 521 (2001) [hep-ph/0108059].\\
G. Degrassi, S. Heinemeyer, W. Hollik, P. Slavich and G. Weiglein,
 Eur. Phys. J. C {\bf 28}, 133 (2003) [hep-ph/0212020].\\
M. Frank, S. Heinemeyer, W. Hollik and G. Weiglein, hep-ph/0212037.\\
S. Heinemeyer, Int. J. Mod. Phys. A {\bf 21}, 2659 (2006) [hep-ph/0407244].\\
S. Heinemeyer, W. Hollik, H. Rzehak and G. Weiglein, Eur. Phys. J. C {\bf 39}, 465 (2005) [hep-ph/0411114].\\
W. Hollik and D. Stockinger, Phys. Lett. B {\bf 634}, 63 (2006) [hep-ph/0509298].\\
M. Frank, T. Hahn, S. Heinemeyer, W. Hollik, H. Rzehak and G. Weiglein, JHEP {\bf 0702}, 047 (2007)[hep-ph/0611326].\\
T.~Hahn, S.~Heinemeyer, W.~Hollik, H.~Rzehak, and G.~Weiglein,  Comput.
Phys. Commun. {\bf 180}, 1426 (2009).
http://www.feynhiggs.de/ .
\bibitem{Martin} S.~P.~Martin, Phys.Rev.D~{\bf 75}:055005, (2007) [arXiv:hep-ph/0701051].
\bibitem{[41]} S.~P.~Martin, Phys. Rev. D {\bf 66}, 096001 (2002) [hep-ph/0206136]; Phys. Rev. D {\bf 65}, 116003 (2002) [hep-ph/0111209]; Phys. Rev. D {\bf 67}, 095012 (2003) [hep-ph/0211366]; Phys. Rev. D {\bf 70}, 016005 (2004) [hep-ph/0312092];
 Phys. Rev. D {\bf 71}, 016012 (2005) [hep-ph/0405022].
\bibitem{[45]} A. Dedes, G. Degrassi and P. Slavich, Nucl. Phys. B 672,
144 (2003) [hep-ph/0305127].
\bibitem{[46]} J.S. Lee et al, Comput. Phys. Commun. 156, 283 (2004)
[hep-ph/0307377].
\bibitem{[49]} B.C. Allanach, A. Djouadi, J.L. Kneur, W. Porod and
P. Slavich, JHEP 0409, 044 (2004) [hep-ph/0406166].
\bibitem{Kant1} R.~V.~Harlander, P.~Kant, L.~Mihaila and M.~Steinhauser,
Phys.\ Rev.\ Lett.\  {\bf 100}, 191602 (2008) [{\bf 101} (2008) 039901]
[arXiv:0803.0672 [hep-ph]].
\bibitem{Kant} P.~Kant, R.~V.~Harlander, L.~Mihaila, and M.~Steinhauser,
JHEP {\bf 1008}, 104 (2010) [arXiv:1005.5709v2 [hep-ph]].\\
http://www-ttp.particle.uni-karlsruhe.de/Progdata/ttp10/ttp10-23/ .
%
\bibitem{HHH} H. Haber, R. Hempfling, and A. H. Hoang, Z. Phys. C {\bf 75} (1997) 539.
%
\bibitem{Carena} M.~Carena, M.~Quiros, and C.~E.~M.~Wagner, Nucl.~Phys.~B{\bf 461}, 407 (1996).
\bibitem{Carena1} M.~Carena, P.~H.~Chanowski, S.~Pokorski, and C.~E.~M.~Wagner,
Phys.~Lett.~B{\bf 441}, 205 (1998).
\bibitem{Heinemeyer} S. Heinemeyer, W. Hollik and G. Weiglein, Phys. Rev. D
58, 091701 (1998) [hep-ph/9803277]; Phys. Lett. B 440,
296 (1998) [hep-ph/9807423].
\bibitem{Hollik} M.~S.~Carena, H.~E.~Haber, S.~Heinemeyer, W.~Hollik, C.~E.~M.~Wagner, and G. Weiglein, Nucl.~Phys.~B{\bf 580}, 29 (2000); hep-ph/0001002
%
\bibitem{Zhang:1998bm}
 R.~-J.~Zhang,
Phys.\ Lett.\ B {\bf 447} (1999) 89  [hep-ph/9808299].  
%
\bibitem{mtb} V.~Barger, M.~S.~Berger, and P.~Ohmann, Phys.~Rev.~D{\bf 47}, 1093 (1993).\\
V.~Barger, M.~S.~Berger, P.~Ohmann, and R.~J.~N.~Phillips, Phys.~Lett.~B{\bf 314}, 351 (1993).
\bibitem{mtb2} U.~Langenfeld, S.-O.~Moch, and P.~Uwer,  Nuovo Cim. C033N4 (2010) 47 
[arXiv:1006.0097v1[hep-ph]].
\bibitem{mtb3} A.~L.~Kataev and V.~T.~Kim, Phys.Part.Nucl. 41 (2010) 946 
[arXiv:1001.4207v2[hep-ph]].
%
\bibitem{BKK} F.~Brummer, S.~Kraml, and S.~Kulkarni,  JHEP 1208 (2012) 089 
[arXiv:1204.5977v2[hep-ph]].
\bibitem{CCB} J.~M.~Frere, D.~R.~T.~Jones, and S.~Raby, Nucl.~Phys.~B{\bf 222}, 11(1983).
\bibitem{CCB1} J.~P.~Derendinger and C.~A.~Savoy, Nucl.~Phys.~B{\bf 237}, 307(1984).
\bibitem{CCB2} C.~Kounnas, A.~B.~Lahanas, D.~V.~Nanopoulos, and wM.~Quiros, Nucl.~Phys.~B{\bf 236}, 438(1983).  See, Appendix C.
%
\bibitem{BB2} H.~Baer, V.~Barger, P. Huang, and A.~Mustafayev, Phys.Rev. D84 (2011) 091701 
[arXiv:1109.3197v2[hep-ph]].
\bibitem{ISASUSY} ISAJET by H. Baer, F. Paige, S.  Protopopescu and
X. Tata, hep-ph/0312045M.
\bibitem{ISASUSYA}
H. Baer, J. Ferrandis, S. Kraml and
W. Porod, Phys. Rev. ~D{\bf 73}, (2006) 0115110.
%
\bibitem{softsusy} B.C.~Allanach, Phys.~Commun.~{\bf143}~(202)305-331.
\bibitem{spheno}W.~Porod, Comput.~ Phys.~Commun. {\bf153}(2003)275
\bibitem{spheno2}W.~Porod and F.~Staub. arXiv:1104.1573.
\bibitem{suspect}A.~Djouadi, J-L.~Kneur, G.~Moultaka.  Comput.~ Phys.~Commun.{\bf176}(2007)426-455.
%
\bibitem{Stal} S.~Heinemeyer, O.~Stal, and G.~Weiglein, Phys.~Lett.~B{\bf 710}, 201 (2012)[arXiv:1112.3026[hep-ph]].
\bibitem{Heng} J.~Cao, Z.~Heng, J.~M.~Yang, Y.~Zhang, J.~Zhu, JHEP 1203 (2012) 086 
[arXiv:1202.5821[hep-ph]].
\bibitem{Badziak} M.~Badziak, E.~Dudas, M.~Olechowski, and S.~Pokorski, JHEP 1207 (2012) 155 
[arXiv:1205.1675[hep-ph]].
%
\end{thebibliography}

\end{document}